# Enhanced Structural Stability and Photo Responsiveness of $CH_3NH_3SnI_3$ Perovskite via Pressure-Induced Amorphization and Recrystallization

*Xujie Lü, Yonggang Wang, Constantinos C. Stoumpos, Qingyang Hu, Xiaofeng Guo, Haijie Chen, Liuxiang Yang, Jesse S. Smith, Wenge Yang, Yusheng Zhao, Hongwu Xu, Mercouri G. Kanatzidis, and Quanxi Jia*

[*]     Dr. X. Lü, Dr. Q. X. Jia
Center for Integrated Nanotechnologies, Los Alamos National Laboratory
Los Alamos, NM 87545, USA
E-mail: xujie@lanl.gov, qxjia@lanl.gov
        Dr. X. Lü, Dr. X. Guo, Dr. H. Xu
Earth and Environmental Sciences Division, Los Alamos National Laboratory
Los Alamos, NM 87545, USA
E-mail: hxu@lanl.gov
        Dr. Y. Wang, Prof. Y. Zhao
High Pressure Science and Engineering Center (HiPSEC), University of Nevada Las Vegas
Las Vegas, NV 89154, USA
        Dr. C. C. Stoumpos, Dr. H. Chen, Prof. M. G. Kanatzidis
Department of Chemistry, Northwestern University
Evanston, IL 60208, USA
E-mail: m-kanatzidis@northwestern.edu
        Dr. Y. Wang, Dr. L. Yang, Dr. W. Yang
High Pressure Synergetic Consortium (HPSynC), Carnegie Institution of Washington
Argonne, IL 60439, USA
        Dr. Qingyang Hu
Geophysical Laboratory, Carnegie Institution of Washington, Washington, DC 20015, USA
        Dr. J. S. Smith
High Pressure Collaborative Access Team (HPCAT), Carnegie Institution of Washington
Argonne, IL 60439, USA
        Dr. Qingyang Hu, Dr. W. Yang
Center for High Pressure Science and Technology Advanced Research (HPSTAR)
Shanghai 201203, China

Perovskite solar cells (PSCs) based on hybrid organic-inorganic halide perovskites have garnered great attention because of their potentially high energy conversion efficiency and relatively low processing cost.[1-4] Since the first efficient solid-state perovskite solar cell with a power conversion efficiency (PCE) of 9.7% was reported in middle 2012, unprecedentedly rapid progress has been made, achieving the PCEs over 20% recently.[5-10] However, these high PCE values are often reached with low stability, which is mainly associated with the structural instability of hybrid perovskites.[1,11] Among the organometal halide perovskites, methylammonium lead triiodide ($CH_3NH_3PbI_3$) together with its mixed halides $CH_3NH_3PbI_{3-x}Br_x$, were mostly studied and exhibited excellent photovoltaic properties because of their strong light absorption and high carrier mobility. Yet the use of lead as a component in these compounds brings toxicity issues and raises environmental concerns during device fabrication, deployment and disposal. To avoid this problem, lead-free organotin halide perovskites (*e.g.* $CH_3NH_3SnI_3$) have been synthesized and employed as light absorbers.[12-14] These recent achievements have been considered as a step forward for the realization of low-cost, high-efficiency and environmental friendly PSCs.

To date, various chemical methods have developed to modify the structural and optoelectronic properties of perovskite absorbers.[15-17] These efforts have focused on optimizing the chemical compositions and increasing their crystallinity via halide mixing, hetero-elemental combination, crystal growth control, *etc*. In addition, different processing approaches such as one-step and sequential solution deposition,[7,18-19] vacuum evaporation,[6] solvent engineering,[20] and vapor-assisted solution processing,[21] have been developed to improve the absorber properties and consequent device performance. These chemical and processing modifications have shown great potential in improving the photovoltaic properties of this class of light harvesting materials towards higher PCEs, yet the stability issues remain. For the modifications on structural and optoelectronic properties, chemical pressure induced by lattice mismatch, together with synthetic temperature, has proved to play critical roles.[22] An alternative to chemical pressure is external pressure, as it can provide a direct way to adjust interatomic distance and hence effectively tune the crystal structure and electronic properties.

In recent years, high pressure techniques have been widely used to modify the physical and chemical properties of various materials and to further our understanding of structure-property relationships.[23,24] Moreover, high-pressure research enables the development of novel materials with emergent or enhanced properties, which otherwise cannot be achieved using traditional techniques.[25-27] Thus far, a few studies have been conducted on structural evolution of organometal halide perovskites under high pressure,[28-30] and revealed a amorphization-recrystallization phenomenon. However, no research has focused on the differentiation of properties between the original sample and high-pressure treated product. In this work, we for the first time compare the structural stability, electrical conductivity, and photo responsiveness of a lead-free tin halide perovskite, $CH_3NH_3SnI_3$, before and after high-pressure treatments up to 30 GPa. The pressure-driven phase transitions and evolution of associated electrical and optoelectronic properties have been characterized by in situ synchrotron X-ray diffraction (XRD), Raman spectroscopy, electrical resistance and photocurrent measurements during two sequential compression and decompression cycles. Our findings reveal improved structural stability, increased electrical conductivity, and enhanced photo responsiveness for $CH_3NH_3SnI_3$ via a pressure-induced amorphization and recrystallization process. Hence, this work not only provides the first comparative study on the structural stability and optoelectronic properties of $CH_3NH_3SnI_3$ before and after its high-pressure treatments, but, more broadly, also opens up a new perspective on understanding the fundamental relationship between local structures and electronic properties of organic-inorganic hybrid perovskites.

The organic-inorganic perovskites can be denoted by the chemical formula, $AMX_3$, where A is an organic cation (such as $CH_3NH_3^+$), M is a metal cation (such as $Sn^{2+}$), and X is a halide anion (such as $I^-$). Similar to the well-known inorganic oxide perovskites,[31] their crystallographic stability and probable structure can be deduced using the tolerance factor, $t = (R_A + R_X)/[\sqrt{2}(R_M + R_X)]$, which is defined as the ratio of the A–X distance to the M–X distance in an idealized solid-sphere model ($R_A$, $R_M$ and $R_X$ are the ionic radii of A, M and X, respectively). For halide perovskites, if $t$ lies in the range 0.89–1.0, the cubic structure is likely, while lower $t$ values give less symmetric tetragonal or orthorhombic structures.[32] Given that the ionic radii of $CH_3NH_3^+$, $Sn^{2+}$ and $I^-$ are, respectively, 1.8 Å, 1.1 Å and 2.2 Å, the tolerance factor $t$ of $CH_3NH_3SnI_3$ (hereafter $MASnI_3$) is calculated to be 0.86.

Experimentally, XRD structural analysis reveals that $MASnI_3$ adopts a tetragonal symmetry with the space group *P4mm* at ambient conditions (Figure 1a).[12,13] The corner-sharing $[SnI_6]^{4-}$ octahedra form an infinite three-dimensional (3D) framework with Sn–I–Sn bond angles of 177.4° and 180.0° along the *a*- and *c*-axes, respectively. This deviation from the ideal cubic structure (*Pm–3m*) arises from polarization of $CH_3NH_3^+$ along the C–N bond direction (parallel to the *c* axis), which slightly distorts the 3D $[SnI_3]^-$ framework and results in a tetragonal (pseudo-cubic) structure. It is well known that structural transitions in perovskites between various crystallographic symmetries on heating or cooling are common, with the high-temperature phase generally being higher symmetry. In addition to temperature, pressure is another state parameter which provides an effective way to adjust interatomic distance and hence affects the crystallographic symmetry.

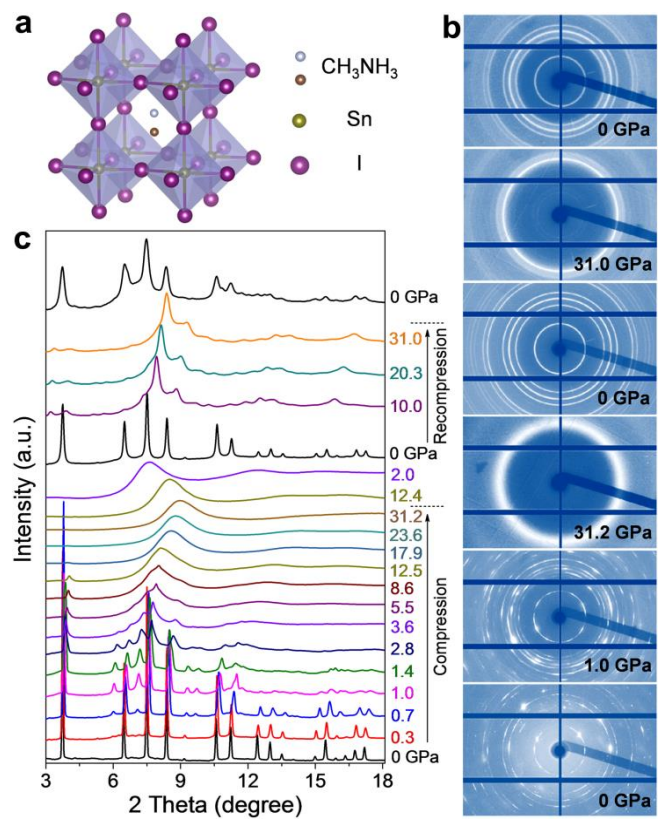

**Figure 1.** In situ structural characterization of MASnI$_3$ under high pressures. (a) Crystal structure of MASnI$_3$. (b) 2D synchrotron XRD images at six selected pressures and (c) integrated 1D XRD patterns of MASnI$_3$ during two sequential compression−decompression cycles.

The MASnI$_3$ sample was loaded in a symmetrical diamond anvil cell (DAC) for in-situ high pressure measurements up to ~30 GPa (see experimental details in the Supporting Information). Figure 1b shows the two-dimensional (2D) XRD images at six selected pressures and Figure 1c shows the integrated 1D XRD profiles during two sequential compression−decompression cycles. With increasing pressure during the 1$^{st}$ compression process, the Bragg diffraction peaks weakened gradually and new peaks appeared at 0.7 GPa, indicating a pressure-induced phase transition. When the applied pressure exceeded 3 GPa, significant degrees of structural disorder occurred, as evidenced by the appearance of a broad diffuse background and the disappearance of some Bragg diffraction peaks, implying the onset of partial amorphization. As the pressure was further increased to 12.5 GPa, all Bragg diffraction peaks disappeared and three broad peaks (one strong and two weak) associated with an amorphous phase were observed. Upon decompression, the amorphous phase was recrystallized to a perovskite structure below 2.0 GPa (more detailed information can be found in Figure S1 in the Supporting Information).

The Rietveld analysis profiles for representative XRD data collected at 1 atm, 0.7 GPa, and after decompression are shown in Figure S2. A tetragonal structure with the space group *P4mm* was used to fit the 1 atm pattern, yielding lattice parameters $a = 6.240(1)$ Å, $c = 6.227(2)$ Å and $V = 242.44(8)$ Å$^3$; while an orthorhombic structure with the space group *Pnma* was used to fit the pattern collected at 0.7 GPa (and those at higher pressures), giving lattice constants $a = 12.348(3)$ Å, $b = 12.300(3)$ Å, $c = 12.355(5)$ Å and $V = 1876.5(9)$ Å$^3$. In hybrid perovskites, the degree of octahedral tilting increases with increasing pressure because the extra-framework MA cations are more compressible than the SnI$_6$ octahedra.[33] Hence, the tetragonal-to-orthorhombic transformation arises mainly from the tilting of SnI$_6$ octahedra coupled with the deformation and movement of CH$_3$NH$_3^+$ cations on compression. This also leads to the contraction of the 3D framework of SnI$_6$ octahedra (unit-cell volume of 1876.5 Å$^3$ for the orthorhombic phase, compared with 1939.52 Å$^3$ for the tetragonal phase in the same cell setting). For fitting of the data collected after decompression, the *P4mm* structure model was again used, yielding cell parameters $a = 6.211(2)$ Å, $c = 6.211(2)$ Å and $V = 239.6(1)$ Å$^3$. Note that the *a* and *c* values of the decompressed phase are the same within uncertainties, indicating increased degree of its cubicity compared with that of the original perovskite structure. Thus we used the cubic model *Pm–3m* to further fit the pattern of decompressed sample, showing the same cell parameter $a = 6.2111(8)$ Å with a smaller uncertainty. In the *Pm–3m* structure, all the Sn–I–Sn bond angles are 180º, which are different from those in the original *P4mm* structure having bond angles of 177.4º and 180.0º along the *a*- and *c*-axes, respectively. We note that the diffraction peaks of the decompressed sample are broadened compared with those of the initial sample (Figure S2), which is likely due to a combined effect of the size reduction of provskite grains and increased local stains between them after compression. Moreover, the XRD background of the decompressed sample displays a broad hump in the 2θ range of 5–9º, suggesting the possible existence of some remaining amorphous component.

The variations of lattice parameters and cell volume of MASnI$_3$ under high pressure are displayed in Figure 2. A phase transition from the tetragonal *P4mm* (phase I) to orthorhombic *Pnma* (phase II) is evidenced by the appearance of additional weak peaks (compare Figures S2a and S2b) due to doubling of each unit cell axis. We note that orthorhombic MASnI$_3$ shows large anisotropy of axial compressibility; the *c* parameter decreases about twice more rapidly than *a* and *b* with increasing pressure (Figure 2a). This behavior may be attributed to different effects of dumbbell-shape MA molecules on the in-phase and out-of-phase titling of [$SnI_6$]$^{4-}$ octahedra.[34] Fitting cell volume data to the Birch-Murnaghan equation of state ($P(V) = \frac{3K_0}{2}\left[\left(\frac{V_0}{V}\right)^{\frac{7}{3}} - \left(\frac{V_0}{V}\right)^{\frac{5}{3}}\right]\left\{1 + \frac{3}{4}(K' - 4)\left[\left(\frac{V_0}{V}\right)^{\frac{2}{3}} - 1\right]\right\}$) yielded a bulk modulus ($K_0$) of 12.3 GPa with K' being fixed at 4, where $V_0$ is the initial volume, V is the deformed volume, and K' is the derivative of the bulk modulus with respect to pressure.[35] The $K_0$ value is much smaller than those of inorganic perovskites like $SrTiO_3$ with $K_0$=172 GPa,[36] thereby indicating the highly compressible nature of hybrid perovskites. When the pressure exceeds a certain threshold, the perovskite structure cannot hold anymore, resulting in amorphization. Interestingly, upon decompression, the amorphous phase reverts back to the crystalline phase. This memory effect may be ascribed to the elasticity of the organic-inorganic framework, which acts as a "template" for the recrystallization process.[29]

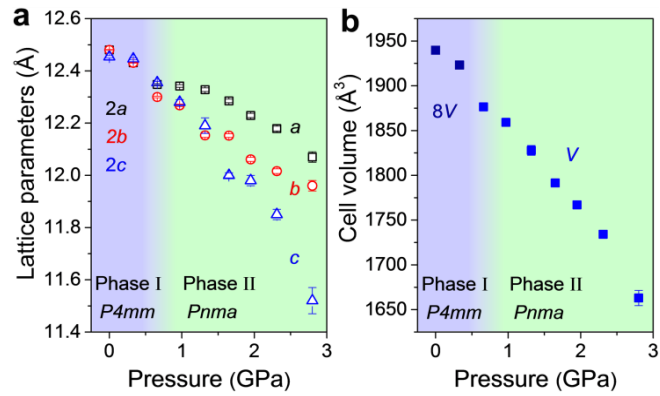

**Figure 2.** Variations of crystal parameters of $MASnI_3$ as a function of pressure. (a) Lattice constants *a* (square), *b* (circle), and *c* (triangle). (b) Cell volume evolution with pressure. For ease of comparison, all cell parameters are plotted on the same scale.

It is worth noting that the decompressed sample shows much smoother diffraction rings (Figure 1b), implying more uniform crystalline grains of $MASnI_3$ after the pressure-induced amorphization and subsequent recrystallization. To further evaluate the differences between the original and pressure-treated $MASnI_3$, we performed the 2$^{nd}$ compression-decompression cycle while simultaneously collecting XRD data. Intriguingly, during the 2$^{nd}$ compression no amorphization was observed up to 31 GPa (Figure 1), though the XRD peaks of the 2$^{nd}$ recovered sample are broader than those of the sample recovered after the 1$^{st}$ compression. It can thus be seen that the pressure-treated sample possesses enhanced structural stability even though it remains the similar perovskite phase. Such an exciting finding motivated us to evaluate the changes of its electrical and photovoltaic properties, as described below.

Electron transport property is one of the most important parameters for photovoltaic materials. To determine its evolution with pressure, we carried out in situ resistance measurements in a DAC during two sequential compression−decompression cycles (the MASnI$_3$ powders were pre-compressed at about 10 MPa to form a dense pellet). The resistance was calculated by Van de Pauw method with the equation, $\exp(-\pi R_1/R_S) + \exp(-\pi R_2/R_S) = 1$, where $R_1$ and $R_2$ are the two resistances measured by the four-probe method, and $R_S$ is the sheet resistance (see details in the Supporting Information). Using the thickness $t$ of 30 μm, the electrical resistivity was calculated as $\rho = R_S \times t$. As shown in Figs 3a and 3b, the resistivity of MASnI$_3$ first decreases with increasing pressure which is typically due to the broadening of the valence and conduction bands, caused by the shortening of bonds.[24,37] Then the resistivity increases sharply until reaching a peak value around 12 GPa, which corresponds to the pressure-induced phase transition and amorphization. Thus there are two competing mechanisms influencing the electron transport: the broadening of valence and conduction bands which plays a dominant role at lower pressures; and the pressure-induced amorphization which is responsible for the precipitous resistance increase at higher pressures. The maximum resistivity is ~6 orders of magnitude higher than the starting value.

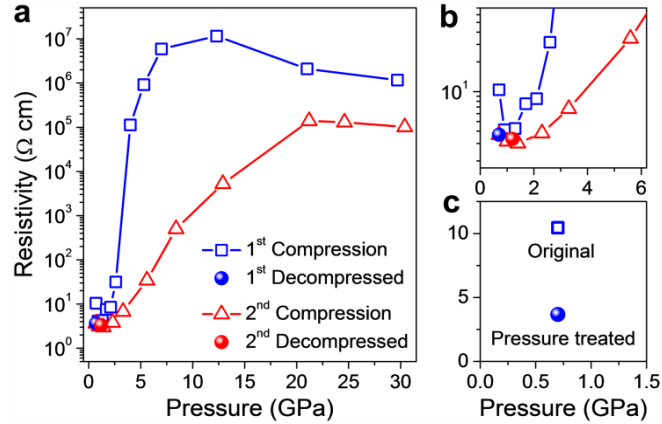

**Figure 3.** In situ resistivity measurements under high pressure up to 31 GPa. (a) Pressure-induced resistivity evolution of MASnI$_3$ during two sequential compression and decompression cycles, blue line shows the first cycle and red line shows the second cycle. Solid spheres represent the values of decompressed phases. (b) Enlarged plot at the lower pressure region. (c) Comparison between the resistivities of MASnI$_3$ before (open square) and after (solid sphere) high-pressure treatments.

During the 2$^{nd}$ compression cycle, the resistivity evolution behavior is somewhat different from that in the 1$^{st}$ cycle. At the beginning, the resistivity decrease is associated with the band broadening — similar to the 1$^{st}$ cycle. However, the subsequent resistivity increase is much slower than that in the 1st cycle corresponding to the pressure-induced amorphization. The resistivity at ~12 GPa for the 2nd cycle is over 3 orders of magnitude lower than that for the 1st cycle, and the maximum resistivity value is also more than 10 times smaller. The improved electron transport ability of the pressure-treated MASnI$_3$ is likely correlated to its enhanced structural stability under high pressure, where no amorphization occurs up to 31 GPa (Figure 1c). These results provide further evidence that the sharp resistivity increase between 3–12 GPa in the 1st compression is due to pressure-induced amorphization. More importantly, the electrical conductivity of the recovered crystalline perovskite from high-pressure treatments is 3 times higher than that of the pristine MASnI$_3$ (Figure 3c). Such enhancement could be partially ascribed to higher carrier mobility. As described earlier, the pressure-treated MASnI$_3$ possesses a higher crystallographic symmetry (cubic) and a smaller unit-cell volume. The cubic symmetry has less anisotropy and broader conduction and valence bands because of the increased orbital overlap (all the Sn–I–Sn bond angles are 180°), which are favorable for higher mobility.[38] Meanwhile, the smaller cell volume indicates shorter Sn–I bonds, which also results in better orbital overlap and broader conduction and valence bands, suggesting smaller carrier effective masses ($m^*$).[38] In order to support this statement, first-principles calculations were performed via density functional theory (DFT) to deteimine the carrier effective masses of MASnI$_3$ perovskite before and after pressure treatments. As shown in Table S1&S2, the calculated $m^*$ of the original MASnI$_3$ (*P4mm*) is 0.083$m_0$ (where $m_0$ is the free electron mass), which is similar to the recently reported values for organic-inorganic hybrid perovskites.[39,40] While $m^*$ of the pressure-treated sample with cubic structure (*Pm–3m*) is reduced to 0.070$m_0$. This smaller effective mass implies higher carrier mobility $\mu$, since $\mu$ is proportional to $1/m^*$.[41] Thus, the MASnI$_3$ after high pressure treatment possesses higher carrier mobility, resulting in increased electrical conductivity. Computational methods and detailed discussions can be found in the Supporting Information. On the other hand, considering the microstructural complexity of a polycrystalline material, some other factors such as grain size/shape, grain boundaries, and the bulk density of the material may also affect the electrical conductivity of pressure-treated MASnI$_3$. In particular, pressure tends to densify a material and it can also induce texture or preferred orientation of the grains.[42,43] As indicated by the changes in the relative XRD peak intensities for the MASnI$_3$ before and after pressure treatments (Figure S2), preferred orientation appeared to occur after compression. These microstructural changes likely contribute to the observed pressure-induced enhancement in conductivity.

Another key parameter for photovoltaic materials is the photo responsiveness which is related to light harvesting ability for solar energy conversion applications. Considering the improved structural stability and increased carrier mobility of the pressure-treated $MASnI_3$, it is not surprising that it may have enhanced photo responsiveness as well. To verify this, we conducted in situ photocurrent measurements under high pressure and compared the photo responsiveness of $MASnI_3$ before and after high-pressure treatments (see details in the Supporting Information). As shown in Figs 4a and 4b, $MASnI_3$ exhibits obvious response to visible light with the on−off switch in the pressure range measured. Despite a large reduction in photocurrent, amorphous $MASnI_3$ still shows discernible values at high pressures, indicating its semiconductor feature as a photovoltaic material. Figure 4a shows the comparison of photocurrents between the original and pressure-treated $MASnI_3$ at a low pressure of 0.7 GPa, and Figure 4b shows their comparison at a high pressure of 25 GPa. Excitingly, the pressure-treated material (after the 1$^{st}$ cycle) exhibits significantly higher photocurrent than the original phase in both low and high pressure regions. Furthermore, the higher dark currents indicate the higher mobilities in the treated $MASnI_3$, consistent with the results from resistance measurements and the DFT calculations.

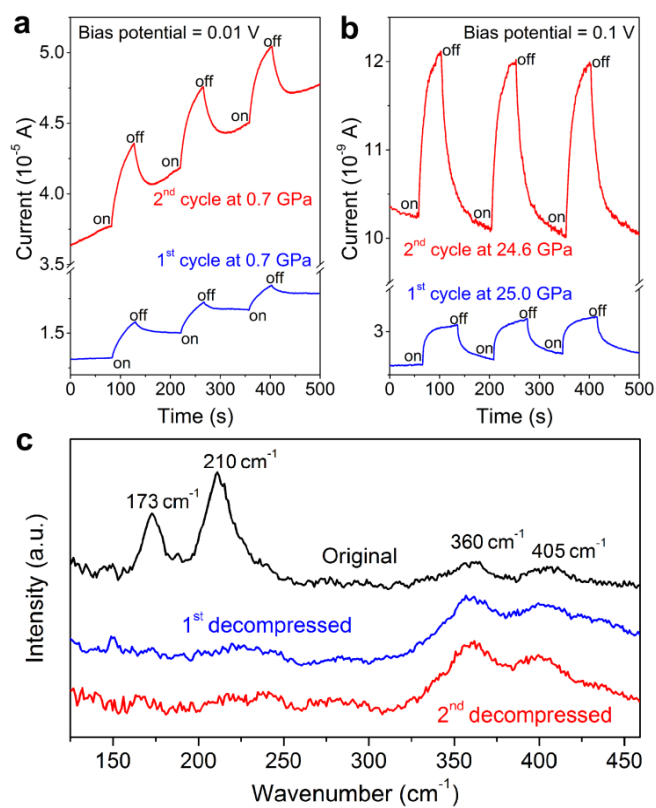

**Figure 4.** Photocurrents of $MASnI_3$ before (1$^{st}$ cycle) and after (2$^{nd}$ cycle) pressure treatments, (a) at a low pressure of 0.7 GPa and (b) at a high pressure of 25 GPa. (c) Raman spectra of the original (black), 1$^{st}$ (blue) and 2$^{nd}$ (red) decompressed $MASnI_3$.

We noticed that the Bragg diffraction peaks of MASnI$_3$ after high-pressure treatments are much broader than those of the original sample (Figure 1c), with their full widths at half maximum (FWHMs) being 0.12° and 0.06°, respectively. This indicates a finer average grain size of the MASnI$_3$ after the pressure-induced amorphization and recrystallization process, which was further verified by the SEM images of MASnI$_3$ before and after high-pressure treatments (Figure S3). Hence, the enhanced structural stability of crystalline hybrid perovskite seems to be associated with a reduction in grain size, suppressing the amorphization above 31 GPa. This behavior follows the so-called Hall-Petch relationship that has been observed in a wide range of materials systems such as TiO$_2$ and cubic BN.[44-46] That is, the grain boundaries act as pinning points impeding propagation of structural disordering under pressure. The Hall-Petch strengthening is a well-known phenomenon and it is not surprising that it also occurs in the hybrid perovskite system. Strikingly, such pressure-treated MASnI$_3$ perovskite exhibits enhanced structural stability, increased electrical conductivity and improved photo responsiveness.

Furthermore, in order to probe possible differences in local bonding (especially for the MA cation), we compared the Raman spectra of the original MASnI$_3$ and the pressure-treated product. As shown in Figure 4c, they present very different Raman features. Using the results of MAPbI$_3$ as reference, the bands at 173 and 210 cm$^{-1}$ can be assigned to librational motions of MA cations.[47,48] These bands disappeared after the high pressure treatments, indicating big changes in local bonding of MA, even though the crystal perovskite structure kept similar (see XRD in Figure 1). The broad 300−450 cm$^{-1}$ features can be assigned to the torsional modes of the MA cations and they are associated to the orientational ordering of MA cations.[47] These torsional modes have no obvious change before and after pressure treatments, which indicates the similar orientation of the organic cations and thus the similar long-range crystal structure. The accurate assignment of the Raman modes in MASnI$_3$ is an important topic but is beyond the scope of this work. The Raman results do provide evidence for the distinctions of local structure in the MASnI$_3$ before and after pressure treatments, which might affect the structural stability and the electronic properties of the hybrid perovskite. As a clean and effective means to tune the atomic lattice and electronic properties of hybrid perovskites, high pressure provides a novel perspective on understanding the fundamental relationship between local structure and physical properties. Then one may optimize the performance of these materials by exploring alternative approaches, *e.g.* via interfacial engineering, to generate/tune local strains for simulating the effects of pressure and/or tension in hybrid perovskite to achieve desired properties at ambient conditions.

In summary, we have reported a comparative study on the structural stability, electrical conductivity, and photo responsiveness of a lead-free tin halide perovskite ($CH_3NH_3SnI_3$) before and after high-pressure treatments. In situ synchrotron XRD, Raman spectroscopy, electrical resistance and photocurrent measurements were conducted on $CH_3NH_3SnI_3$ during two sequential compression−decompression cycles up to 30 GPa. A phase transition from tetragonal to orthorhombic at 0.7 GPa was observed, followed by an amorphization starting at about 3 GPa. Surprisingly, no amorphization can be observed during the re-compression process, and the crystalline nature of pressure-treated $CH_3NH_3SnI_3$ persists to at least 30 GPa, indicating an improved structural stability. In situ high-pressure resistance measurements reveal a three-fold increase in electrical conductivity of the pressure-treated $CH_3NH_3SnI_3$ in comparison with the pristine sample. Photocurrent measurements also demonstrate substantial enhancement in visible-light responsiveness of the perovskite after high pressure treatments. The mechanisms underlying the enhanced structural stability and associated property improvements were investigated. Such pressure-treated photovoltaic materials with superior properties of improved structural stability, increased electrical mobility, and enhanced photo responsiveness will inspire scientists to develop high-performance photovoltaic materials under ambient pressure, for instant, by introducing local strain as an alternative of pressure and/or tension in the hybrid perovskite films.

**Supporting Information**

Supporting Information is available from the Wiley Online Library or from the author.

**Acknowledgements**

X.L. acknowledges the J. Robert Oppenheimer Distinguished Fellowship supported by the Laboratory Directed Research and Development Program of Los Alamos National Laboratory. The work at Los Alamos National Laboratory was performed, in part, at the Center for Integrated Nanotechnologies, an Office of Science User Facility operated for the U.S. Department of Energy, Office of Science. The UNLV High Pressure Science and Engineering Center (HiPSEC) is a DOE/NNSA Center of Excellence supported by Cooperative Agreement DE-NA0001982. HPCAT operations are supported by DOE−NNSA under Award DE-NA0001974 and DOE-BES under Award DE-FG02-99ER45775, with partial instrumentation funding by NSF. W. Y. acknowledges the financial support from US DOE-BES X-ray Scattering Core Program under grant number DE-FG02-99ER45775 and NSAF(Grant No.

U1530402). At Northwestern University this work was supported by grant SC0012541 from the U.S. Department of Energy, Office of Science.